\documentclass[fleqn,usenatbib]{mnras}

\usepackage[T1]{fontenc}

\DeclareRobustCommand{\VAN}[3]{#2}
\let\VANthebibliography\thebibliography
\def\thebibliography{\DeclareRobustCommand{\VAN}[3]{##3}\VANthebibliography}

\usepackage{graphicx}	% Including figure files
\usepackage{amsmath}	% Advanced maths commands
\usepackage{amssymb}	% Extra maths symbols
\usepackage{multicol}        % Multi-column entries in tables
\usepackage{bm}		% Bold maths symbols, including upright Greek
\usepackage{pdflscape}	% Landscape pages
\usepackage{cancel}
\usepackage[normalem]{ulem}
\usepackage{hyperref}
\hypersetup{%
    colorlinks=true, 
    citecolor=blue} 
\usepackage{xcolor}
\definecolor{webgreen}{rgb}{0,.5,0}
\definecolor{webbrown}{rgb}{.6,0,0}
\definecolor{purple}{rgb}{0.5,0,.5}

\usepackage[T1]{fontenc}
\usepackage{ae,aecompl}

\title[NS magnetic instability \& elastic shear]{Does elasticity stabilise a magnetic neutron star?%\\
}

\author[P. Bera et al.]{
Prasanta Bera\thanks{E-mail: P.Bera@soton.ac.uk (PB)},
 D. I. Jones,
 Nils Andersson
\\
Mathematical Sciences and STAG Research Centre, University of Southampton, Southampton SO17 1BJ, UK\\
}

\date{Accepted XXX. Received YYY; in original form ZZZ}

\pubyear{2020}

\begin{document}
\maketitle
\label{firstpage}
\pagerange{\pageref{firstpage}--\pageref{lastpage}}

\begin{abstract}
The configuration of the magnetic field in the interior of a neutron star is mostly unknown from observations. Theoretical models of the interior magnetic field geometry tend to be oversimplified to avoid mathematical complexity and tend to be based on axisymmetric barotropic fluid systems. These static magnetic equilibrium configurations have been shown to be unstable on a short time scale against an infinitesimal perturbation. Given this instability, it is relevant to consider how more realistic neutron star physics affects the outcome. In particular, it makes sense to ask if elasticity, which provides an additional restoring force on the perturbations, may stabilise the system. It is well-known that the matter in the neutron star crust forms an ionic crystal. The interactions between the crystallized nuclei can generate shear stress against any applied strain. To incorporate the effect of the crust on the dynamical evolution of the perturbed equilibrium structure, we study the effect of elasticity on the instability of an axisymmetric magnetic star. In particular we determine the critical shear modulus required to prevent magnetic instability and consider the corresponding astrophysical consequences.
\end{abstract}


%%%%%%%%%%%%%%%%%%%%%%%%%%%%%%%%%%%%%%%%%%%%%%%%%%

%%%%%%%%%%%%%%%%% BODY OF PAPER %%%%%%%%%%%%%%%%%%
%\begingroup
%\let\clearpage\relax
%\tableofcontents
%\endgroup
\newpage

\section{Introduction}
Axisymmetric magnetic field configurations in a barotropic star are prone to the so-called Tayler instability \citep{Tayler1973,Markey+Tayler1973}. Starting from a static equilibrium, a perturbation with pure poloidal/toroidal magnetic field grows on the characteristic Alfv\'en time scale, as these equilibrium states are not at the minimum energy configuration \citep{Kruskal+Schwarzschild1954,Shafranov1956,Tayler1957,Flowers+Ruderman1977}. Numerical time evolution studies of magneto-hydrodynamic equations show that a stable magnetic field configuration in a radiative star requires mixed poloidal and toroidal magnetic energy components with comparable strength \citep{Braithwaite+Nordlund2006}, following the hypothesis by \cite{Prendergast1956}. This may resolve the problem for main-sequence stars, as a region of radiative convection may be present in their interior. However, a neutron star, formed following a supernova, does not satisfy the condition to have a convective region due to the rapid cooling just after formation. 

\cite{Braithwaite2009} suggested the possibility of stability of a mixed field configuration within a neutron star with stratified matter based on simulations where mixed field configurations are imposed at the beginning of the time evolution. The feature of stable stratification has been identified as an essential criterion of a magnetic star with a mixed field to exhibit long-duration stability in the numerical evolution  \citep{Mitchell+2015} as well as in analytical \citep{Akgum+2013} and semi-analytical formulations \citep{Herbrik+Kokkotas2017}. However, it appears that a self-consistent calculation of axisymmetric barotropic configurations in static equilibrium can not achieve the required mixed field ratio \citep{Armaza+2014}. Instead, the self-consistent field technique generates configurations having either pure toroidal field or a poloidal field with a subdominant ($<10\%$) toroidal component \citep{Tomimura+Eriguchi2005,Lander+Jones2009}. These axisymmetric configurations remain unstable against a small perturbation and the instability grows on the characteristic Alfv\'en time scale \citep{Lander+Jones2011a,Lander+Jones2011b,Bera+Bhattacharya2017}. The problem of an unstable magnetic field remains, but one should keep in mind that these studies are based on an ideal fluid model with no dissipation, e.g. viscosity or resistivity, This makes sense as the instability time scale is much shorter than the dissipation time scales, but it seems relevant to ask if the problem may be resolved by adding more realistic physics, e.g. a non-uniform thermonuclear reaction rate, electrostatic interaction between matter elements, local turbulent velocity etc. to list a few possibilities. In this paper we take a step in this direction by assessing the impact of \emph{elasticity} on the magnetic instability.

An elastic crust is formed in the outer region of a neutron star as it cools rapidly after its birth \citep{Ruderman1968}. Although the crust contains only a few per cent of the total mass of a star, the conductivity of the crust plays a crucial role in the thermal evolution of the neutron star \citep{Lorenz+1993,Page+2000}. In addition, a crust fracture may cause a sudden change in the rotation speed of an isolated neutron star, known as a pulsar star-quake \citep{Jones2003}. The presence and evolution of the electric current in the crust influences the magnetosphere of the neutron star. It may cause observable features like soft gamma-ray bursts from a strongly magnetized system \citep{Thompson+Duncan1995,Thompson+Duncan1996,Thompson+2017,Karageorgopoulos+2019}. The elastic properties of the crustal element can be obtained from studies of electrostatic interaction in the crystalline matter \citep{Carter+2006,Chamel+Haensel2008,Baiko2015}. Molecular studies of the neutron star crust matter indicate that the crust is adequate to maintain considerable stress \citep{Horowitz+Kadau2009}. This stress may be due to an elastic ``mountain'', or any internal magnetic field \citep{Chugunov+Horowitz2010}.

Any perturbation of the system, which leads to the displacement of the matter from its equilibrium position, generates strain within the crystalline structure, and induces stress opposing the applied strain. The volumetric isotropic component of stress is the pressure. In the equilibrium state, the force due to the pressure balances the gravitational force and the Lorentz force (in a magnetic configuration). The trace-free symmetric part of the stress tensor is shear. The modified electromagnetic interactions between the displaced nuclei in the stellar crust generate shear stress. This may act as a restoring force against the perturbation. Here we consider the effect of shear on the evolution of the perturbation of a magnetic star. In this study, the main assumptions are: i) the star is barotropic and non-rotating, ii) the equilibrium structure is axisymmetric, iii) the Cowling approximation is assumed to be valid, i.e. the perturbed state has a fixed gravitational potential. The basic equations used for the study, are provided in section~\ref{S2}. A brief description of the formalism used to solve these equations is provided in section~\ref{S3}. The results and conclusions are summarized in section~\ref{S4} and section~\ref{S5}, respectively.

\section{Basic equations}\label{S2}
Because of the compactness of a neutron star, a precise model requires a general relativistic formalism for both gravity and the magnetic field. However, as we are providing a proof of principle, we simplify the calculation by considering the neutron star as a fluid system in Newtonian gravity. We use a spherical polar coordinate system $(r, \theta, \phi)$ with its origin at the centre of the star.
\subsection{Magneto-hydrodynamic (MHD) equations}
The dynamical behaviour of the fluid are described by the continuity and Euler equations corresponding to the mass and momentum conservation:
\begin{align}
\frac{\partial\rho}{\partial t} &= -\nabla\cdot(\rho\boldsymbol{v})\\
    \left(\frac{\partial\boldsymbol{v}}{\partial t}+(\boldsymbol{v}\cdot\nabla)\boldsymbol{v}\right) &= -\frac{1}{\rho}\mathbf{\nabla} \boldsymbol{\Pi} -\mathbf{\nabla} \Phi_g+\frac{1}{\rho}\left( \boldsymbol{J}\boldsymbol\times\boldsymbol{B}\right).
\end{align}
Here $\rho$, $\Phi_g$, $\boldsymbol{J}$ and $\boldsymbol{B}$ are the matter density, gravitational potential, electric current density and magnetic field, respectively. $\frac{\partial}{\partial t}$ and $\nabla$ represent time and space derivatives. From the perturbed displacement $\boldsymbol{\xi}$, the matter velocity ($\boldsymbol{v}$) follows as,
\begin{align}\label{xi_eqs}
\frac{\partial\boldsymbol{\xi}}{\partial t} &= \boldsymbol{v}.
\end{align}
The stress tensor $\boldsymbol{\Pi}$ is a combination of a symmetric pressure (isotropic) term $p$ and the anisotropic tensor $\mathbf{T}$, i.e. 
\begin{align}
    \boldsymbol{\Pi}&=p\mathbf{I}+\mathbf{T},
\end{align}
with $\mathbf{I}$ as the unit rank-three tensor. The gravitational potential $\Phi_g$ is related to the matter density by Poisson's equation,
\begin{align}
    \mathbf{\nabla}^2\Phi_g &= 4\pi G \rho,
\end{align}
where $G$ is the gravitational constant. The magnetic field components satisfy the divergence-free condition $\mathbf{\nabla}\cdot \boldsymbol{B} = 0$ and $\mathbf{\nabla}\times \boldsymbol{B} = \mu_0\boldsymbol J$, $\mu_0$ being the free space permeability. In the ideal MHD limit we neglect the effect of resistivity and the evolution equation for the magnetic field becomes 
\begin{align}
\frac{\partial \boldsymbol B}{\partial t} &= \nabla\times(\boldsymbol{v}\times\boldsymbol{B})
\end{align}
To solve the set of equations, we need to provide a relation between matter pressure and density, i.e. $p(\rho)$, which describes the equation of state (EoS). Here we consider a polytropic EoS $p \propto \rho^{1+\frac{1}{n}}$, with $n$ the polytropic index. We consider the presence of elasticity by including the anisotropic tensor $\mathbf{T}$ due to the stress as discussed in \ref{sS2_3}.

\subsection{Perturbed equations}
To study the dynamics of a magnetic neutron star, we consider perturbations of the MHD equations. Initially, we find equilibrium configuration assuming a static system employing the self-consistent field technique \citep{Hachisu1986,Jones+2002,Tomimura+Eriguchi2005,Lander+Jones2009,Bera+Bhattacharya2014,Bera+Bhattacharya2016}. The total magnetic energy ($\mathcal{M}$) and gravitational energy ($\mathcal{W}$) of the configuration are obtained from the volume integrals $\mathcal{M}=\int dV \frac{B^2}{2\mu_0}$ and $\mathcal{W}=\frac{1}{2}\int dV\Phi_g\rho$, $dV$ being volume element. We then consider the linear evolution of the perturbed components with respect to the static equilibrium star. Restricting the analysis to linear order is numerically less expensive and the evolution of the entire star can be done without introducing an artificial atmosphere. This method comes with some limitations, too. For example, we can not infer the final outcome if the perturbations grow by a significant amount, such that non-linear terms become important. However, we can accurately identify the onset of instability in different environments and establish the characteristics as long as the linear terms dominate the evolution.

All variables are expressed as a sum of the equilibrium part (with subscript `0') and the perturbation, i.e, $p = p_0+\delta p$, $\rho = \rho_0+\delta\rho$, $\boldsymbol{B} = \boldsymbol{B_0}+\delta\boldsymbol{B}$. The equilibrium components remain time independent whereas the perturbed components depend on time. Being restricted to small perturbation, we use the Cowling approximation, neglecting the change in the gravitational potential over time i.e. $\Phi_g={\Phi_g}_0$. The linear expansion of time-dependent equations takes the following form  \citep{Lander+Jones2011a,Lander+Jones2011b,Bera+Bhattacharya2017,Bera+Bhattacharya2017c}, 

\begin{align}\label{hydro_pertb}
\frac{\partial\boldsymbol{\xi}}{\partial t} &= \boldsymbol{v}\\
\frac{\partial\boldsymbol{f}}{\partial t} & = \left[-\mathbf{\nabla} \delta p + \frac{\mathbf{\nabla} p_0}{\rho_0}\delta\rho \right] \nonumber\\
& -\frac{1}{\rho_0}\left(\boldsymbol{J_0}\boldsymbol\times\boldsymbol{B_0}\right)\delta\rho +  \frac{1}{\rho_0}\left(\boldsymbol{J_0}\boldsymbol\times\boldsymbol{\beta}\right) \nonumber\\
& + \left[\frac{1}{\rho_0}\left(\nabla\times\boldsymbol{\beta}\right)\times\boldsymbol{B_0} - \frac{1}{\rho_0^2}\left(\nabla\rho_0\times\boldsymbol{\beta}\right)\times\boldsymbol{B_0}\right]\nonumber\\
& + \boldsymbol{\mathcal{F}}_{shear}\label{perturbedEq_with_shear}\\
\frac{\partial\delta\rho}{\partial t} & = -\nabla\cdot\boldsymbol{f}\\
\frac{\partial \boldsymbol\beta}{\partial t} &= \nabla\times(\boldsymbol{f}\times\boldsymbol{B_0})-\frac{\nabla\rho_0}{\rho_0}\times(\boldsymbol{f}\times\boldsymbol{B_0})\\ 
\delta p & = \left.\frac{\partial p}{\partial \rho}\right|_0\delta\rho \label{hydro_pertb_end}
\end{align} 
Here, $\boldsymbol{f}=\rho_0\boldsymbol{v}$ and $\boldsymbol\beta=\rho_0\delta\boldsymbol{B}$. Each of these perturbed quantities is further decomposed in azimuthal angle $\phi$ with index $m$, e.g. the perturbed pressure 
\begin{align}
 &\delta p (t, r, \theta, \phi) = \nonumber\\
 &\sum_{m=0}^{m=+\infty}[\delta p^+(t, r, \theta) \cos m\phi + \delta p^-(t, r, \theta) \sin m\phi].
\end{align}
Hence a specific azimuthal mode, corresponding to the value of $m$, can be evolved in the two-dimensional $r-\theta$ plane considering the axisymetric background star.

The force term due to shear in equation~\ref{perturbedEq_with_shear} is obtained from the anisotropic stress tensor
\begin{align}\label{F_shear}
    \boldsymbol{\mathcal{F}}_{shear} = -\nabla\cdot\textbf{T}.
\end{align}

\subsection{Shear Terms}\label{sS2_3}
We assume that a perturbation is applied at the location $\boldsymbol{x}$ and the deformation displaces it to $\boldsymbol{x}\rightarrow\boldsymbol{x}+\boldsymbol{\xi}(\boldsymbol{x})$. The spatial gradient of the displacement vector measures the strain i.e. 
\begin{align}
    \mathbf{S}=\nabla\boldsymbol{\xi} ~~\textrm{or,}~S_{ik}=\nabla_k\xi_i=\xi_{i;k}=\frac{\partial\xi_i}{\partial x_k}+\Gamma_{ijk}\xi_j.
\end{align}
In a curvilinear coordinate system the unit vectors (e.g. $\textbf{e}_i$) are not fixed and the directional derivatives come with a scaling factor. The connection parameters $\Gamma_{ijk}=\textbf{e}_i\cdot(\nabla_k\textbf{e}_j)$ are used to find the gradients in a curvilinear geometry. 

The trace-less body-shear components of the strain tensor $(\boldsymbol{\Sigma})$ are defined as, 
\begin{align}\label{sigma_ij}
    \Sigma_{ij} = \frac{1}{2}\left(S_{ij}+S_{ji}\right)-\frac{1}{3}g_{ij}S_{kk},
\end{align}
where $g_{ij}$ is the flat metric associated with the coordinates \citep{Thorne+Blandford2017}.
Now applying the Hooke's law of elasticity, we obtain the stress tensor:
\begin{align}
    \mathbf{T}=-2\mu\boldsymbol{\Sigma}.
\end{align}
Here, $\mu$ is the shear modulus. Hence, the shear force term takes the form:
\begin{align}
    \boldsymbol{\mathcal{F}}_{shear} = -\nabla\cdot\textbf{T}=2\mu\nabla\cdot\boldsymbol{\Sigma}+2\nabla\mu\cdot\boldsymbol{\Sigma}.
\end{align}
    
Some of the relevant shear terms in spherical polar coordinates are listed in appendix~\ref{shear_terms}.

\subsection{Instability and shear modulus}\label{sS2_4}
It is known that the Tayler instability is intrinsic to the magnetic field geometry and the temporal characteristics of the instability depend on the effective Alfv\'en time scale or magnetic field strength \citep{Lander+Jones2011a,Bera+Bhattacharya2017}. To compare different configurations, we use the total magnetic energy $\mathcal{M}$ for a specific magnetic field geometry\footnote{In a general situation with different magnetic configurations and/or stellar structures, the ratio between magnetic to gravitational energy ($\mathcal{M/W}$) can be a useful measure of instability strength.}. Here we plan to study the effects of the shear modulus on the magnetic instability. The elasticity of the medium prevents a significant perturbation from its equilibrium. Hence, we desire to find a threshold criterion relating to the presence of stability in a medium with volume integrated shear energy term:
\begin{align}\label{eq_instability_shear}
    \mathcal{E}_{shear}>\kappa\mathcal{M},
\end{align}
with $\kappa$ a parameter depending on the field geometry and elastic properties. Here we have introduced $\mathcal{E}_{shear}$($\equiv\int dV\mu$) as a measure of shear energy corresponding to a system with unit strain. Therefore, for a uniform shear modulus ($\mu$) within the spherical volume of radius $R$, $\mathcal{E}_{shear}=4\pi R^3\mu/3$. It may be noted that the total shear energy of a system under strain is $\int dV\mu\Sigma_{ij}\Sigma_{ij}$.

\section{Stars with no elastic components}\label{S3}

Before considering the effect of including an elastic component in our magnetised star, we will first consider purely magnetic configurations, i.e.\ no elastic component.  The aim is to exhibit the magnetic instabilities described previously, as these  will provide a point of comparison for the results of section~\ref{S4}, where an elastic component is included.

To this end, in Section~\ref{sS3_1} we describe our method for computing equilibrium magnetic field configurations, and in section~\ref{sS3_2} we describe numerical time evolutions of small perturbations of these configurations.

\begin{figure*}%-------------------------------------------------------------------
	\includegraphics[width=\columnwidth]{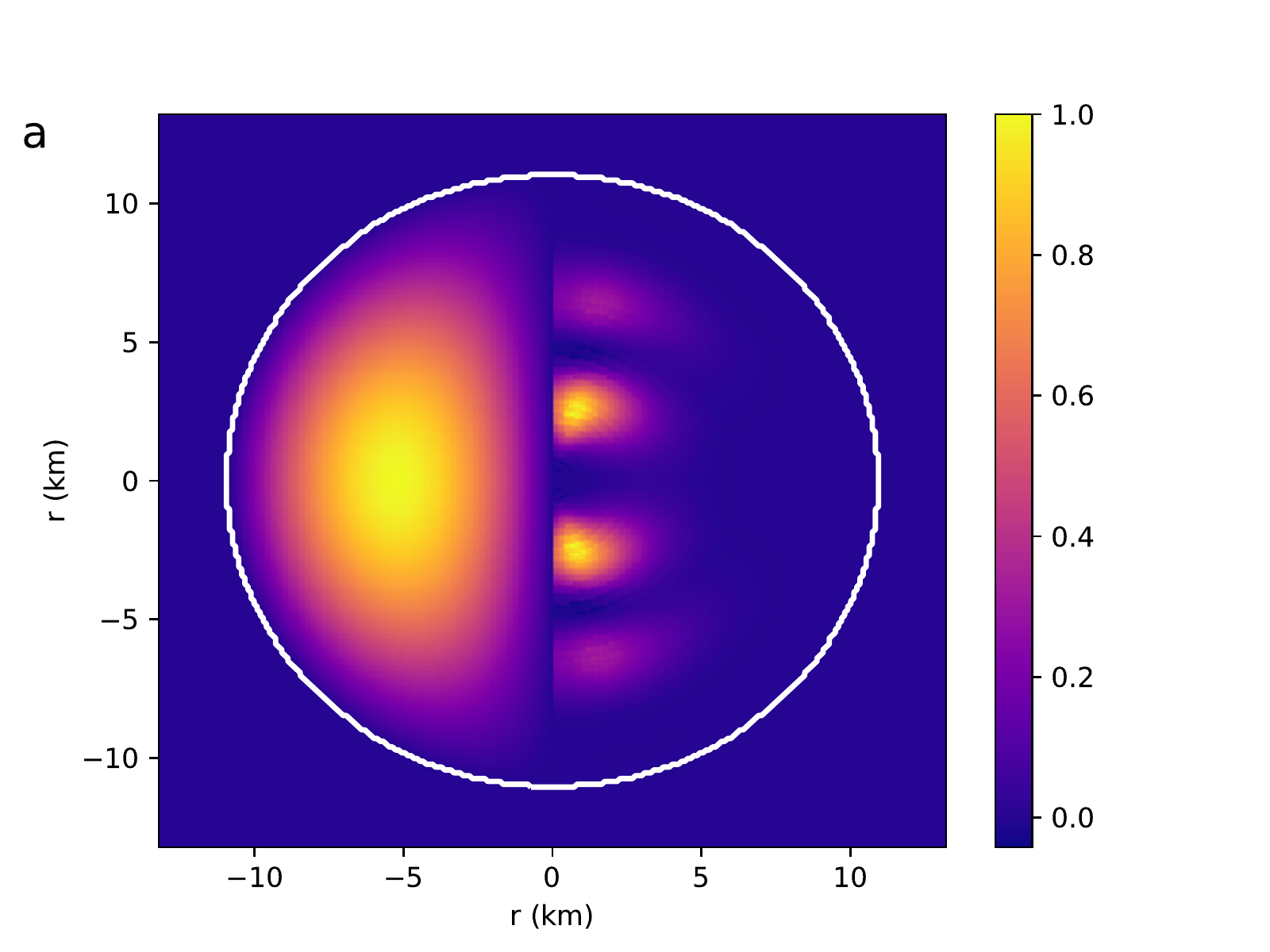}%{B0dBtor.pdf}
	\includegraphics[width=\columnwidth]{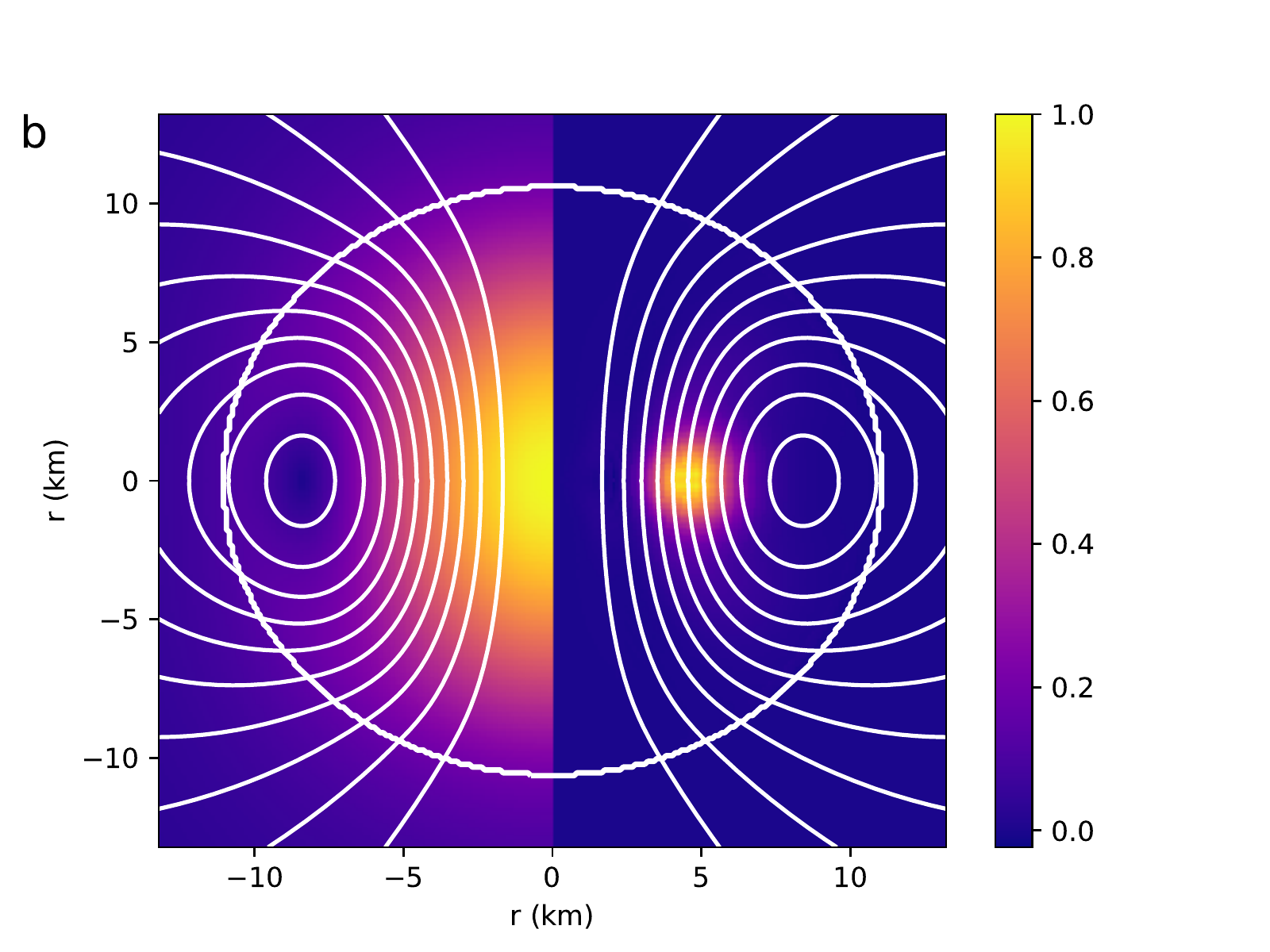}%{B0dBpol.pdf}
    \caption{a) An axisymmetric pure toroidal magnetic field is confined within the stellar surface (white line) of a 1.23~$M_\odot$ neutron star with magnetic to gravitational energy ratio $|\mathcal{M/W}|=0.008$. The magnitude of the toroidal field strength normalised to the maximum, i.e. $|B_0|/|B_0|_{max}$ (left half) and the growth of perturbed magnetic energy ($|\delta \mathcal{M}|$) in some arbitrary unit (right half) from the $l=m=1$ linear order time evolution are shown in the colour scale. The toroidal magnetic field lines are perpendicular to the meridian plane. In the perturbed state, the instability arises close to the azimuthal axis. b) The axisymmetric poloidal magnetic field lines in the meridian plane (in white lines) extend beyond the stellar surface (in thicker white line) of a polytropic star. In this case the average magnetic field strength is about $10^{13}$~T and $|\mathcal{M/W}|=0.008$. The colour scale in the left-half indicates the equilibrium field strength normalised with maximum strength (i.e. $|B_0|/|B_0|_{max}$) occurring at the stellar core. The minimum magnetic field strength is found in the central region with concentric poloidal field lines at a finite radius from the stellar core in the equatorial plane. The other half (right) indicates the perturbed magnetic energy ($\propto|\delta \mathcal{M}|$ in some arbitrary unit) at the onset of instability for an $l=m=2$ perturbation. The most significant perturbed magnetic energy occurs near the half-way region between the magnetic null region and the stellar core on the equatorial plane. 
    }\label{fig1:static}
\end{figure*}

%>---------------------------------------------------------------
\subsection{Equilibrium configurations}\label{sS3_1}
We construct equilibrium axisymmetric configurations of polytropic neutron stars with magnetic fields.  We assume that the equilibrium configurations are elastically relaxed, so there are no elastic forces acting. For a fixed central density and magnetic energy, we iteratively solve the stellar structure equations using the self-consistent field technique \citep{Hachisu1986,Tomimura+Eriguchi2005}. As noted above, a mixed poloidal-toroidal configuration with significant contribution from both components with these assumptions remains unachievable.  We therefore choose to consider only purely toroidal and purely poloidal magnetic field configurations.   The matter and magnetic field distributions we compute satisfy the stellar virial criteria. 

For example, we configure a polytropic ($n=1$) neutron star with mass 1.23~$M_\odot$ and equatorial radius 11 km having a pure toroidal magnetic field such that the magnetic to gravitational energy ratio $|\mathcal{M/W}|= 0.008$ (Figure~\ref{fig1:static}a). The star has a very little non-spherical deformation with polar to equatorial radius ratio $R_p/R_{eq}=1.01$. Here the toroidal magnetic field is confined within the star. The field vanishes at the stellar surface and along the axis of symmetry. The magnitude of the field gradually increases from the centre and reaches a maximum value of about $2\times10^{13}$~T. The region with a strong magnetic field is at an intermediate radial distance towards the surface on the equatorial plane. 

Similarly, we construct a poloidal field configuration with mass 1.27 $M_\odot$, radius 11 km ($R_p/R_{eq}= 0.96$) and $|\mathcal{M/W}|= 0.008$ (Figure~\ref{fig1:static}b). Here the magnetic field is extended beyond the stellar surface. The maximum field strength is about $3\times10^{13}$~T at the core of the star, and it vanishes at the centre of the concentric field lines on the equatorial plane. Hence, for a pure poloidal magnetic field geometry, the magnetic ``null-points" form a ring in the equatorial plane. 

We calculate characteristic time scales, Alfv\'en crossing time ($\tau_A$) and sound crossing time ($\tau_s$) of these configurations using volume-averaged variables:
\begin{align}
    \tau_A&= R\frac{\sqrt{\mu_0\bar{\rho_0}}}{\bar{B_0}},\label{eq_tauA}\\
    \tau_s&= R\sqrt{\frac{\bar{\rho_0}}{\left(1+\frac{1}{n}\right)\bar{p_0}}}.
\end{align}
Here, $\bar{\rho_0}$, $\bar{B_0}$ and $\bar{p_0}$ are volume averaged matter density, magnetic field strength and pressure respectively. The characteristic Alfv\'en crossing time of these configurations are about 1~\rm{ms} whereas the sound crossing time is about two orders of magnitude smaller than that. We consider these configurations with rather strong magnetic fields as the reduced $\tau_A$ requires a shorter computation time to exhibit the phenomenon connected to the magnetic field. 

Note that the general characteristics are similar for the configurations with different field strengths (i.e. different $|\mathcal{M/W}|$). A configuration with a stronger magnetic field has a larger deformation (e.g. a pure poloidal configuration with $|\mathcal{M/W}|=0.016$ has a $R_p/R_{eq}=0.92$). 

%---------------------------------------------------------------<

%>---------------------------------------------------------------

\subsection{Time evolution}\label{sS3_2}
An equilibrium configuration can be perturbed in various ways by changing stellar variables \citep{Lockitch+Friedman1999}. Here we consider an axial velocity perturbation in the form:
\begin{align}\label{ini_v_pert}
    \boldsymbol{v} = f(r)\,\hat{r}\times\nabla Y_{lm}(\theta,\phi)
\end{align}
where, $Y_{lm}(\theta, \phi)$ is the spherical harmonic and $f(r)$ is a function dependent on~$r$.  We will specialise to the case of $l=m$, as was done in several previous studies, e.g. \citet{Lander+Jones2011a, Lander+Jones2011b}.

We evolve the linear order perturbation equations forward in time maintaining the suitable boundary conditions \citep{Lander+Jones2011a}. To achieve a long term evolution avoiding spurious small-scale oscillations, we add a fourth-order Kreiss-Oliger dissipation to the dynamic evolutionary equations~(\ref{hydro_pertb}-\ref{hydro_pertb_end}). We also implement a hyperbolic divergence cleaning method to ensure the minimum effects from numerical error in calculating the magnetic field components \citep{Dedner+2002}. In the following, we present a summary of the relevant characteristics (Figure~\ref{fig1:static}) obtained from the time evolution of the perturbed variables where the shear modulus does not make an impact, i.e. $\mu=0$.

The time evolution of the perturbed variables on the static background of the equilibrium magnetic configuration shows the growth of the kinetic and magnetic energy components. At $t=0$, the initial velocity perturbation drives the flow of the matter, which in turn perturbs the  magnetic field, generating additional Lorentz forces.   The presence of exponential growth in the time evolution of the perturbed magnetic energy shows instability. The characteristics are similar in both field geometries (Figure~\ref{fig1:static}). 

These axisymmetric magnetic field geometries were already known to exhibit magnetic Tayler instability. Different earlier studies indicate that the $m=1$ mode of a star with a toroidal magnetic field configuration is strongly unstable \citep{Tayler1973,Lander+Jones2011a}, which motivates our choice of focusing on the $m=1$ mode in the toroidal case.  For a pure poloidal field geometry, the linear order time evolution studies of the perturbed variables with $m=1,~2,~4$ indicate that as the $m$ value increases, the instability growth rate increases \citep{Lander+Jones2011b}. The higher $m$ modes would suffer from the effect of dissipation, which we do not explicitly model in our study. We therefore choose to evolve the $m=2$ mode for a purely poloidal field geometry. 

The perturbations in the matter and magnetic field influence the characteristics of time evolution via the sound and Alfven waves. In this linear-order approximation study, the speed of these waves depends on the background equilibrium values. The sound wave propagates maximally at the core of the star. On the other hand, the speed of the Alfv\'en wave is maximum where the field strength dominates over the matter density. The combined effect of these waves creates a significant perturbation near the most unstable region. We define the (global) growth rate of instability as a logarithmic time derivative of the perturbed component of magnetic energy:
\begin{align}
\zeta=\frac{\partial\ln(\delta\mathcal{M/M})}{\partial t}.
\end{align}
The average growth rate of the instabilities in Figure~\ref{fig1:static}~a~\&~b are about $5\times10^{3}~\rm{s^{-1}}$. The $m=1$ perturbation mode in toroidal magnetic field configuration has a slightly higher growth rate compared to the $m=2$ mode in poloidal field configuration (Figures~\ref{fig2:evolution_in_shear},~\ref{fig3:rateVSmu},~\ref{fig4P:rateVSmu}).

This linear order time evolution captures the characteristics as long as the magnitude of the perturbation is small compared to the background structure. So, we can consider the results with certainty close to the onset of instability.

\section{Results}\label{S4}
Applying the formalism mentioned in section~\ref{S3}, we now study the evolution of perturbed stellar variables of the axisymmetric magnetic configurations, including the effects of shear modulus.  In the beginning, we start with $\boldsymbol{\xi}=\boldsymbol{0}$ everywhere within the star. At a later time, the time integration of the equation~\ref{xi_eqs} gives the value of $\boldsymbol{\xi}$. The components of the strain tensor ($\Sigma_{ij}$) are calculated following equation~\ref{sigma_ij}. We assume that the components of the strain tensor vanishes at the boundary points.

\subsection{Evolution of stellar magnetic field for a uniform shear modulus}\label{sS4_1}
\begin{figure*}%-------------------------------------------------------------------
	\includegraphics[width=\columnwidth]{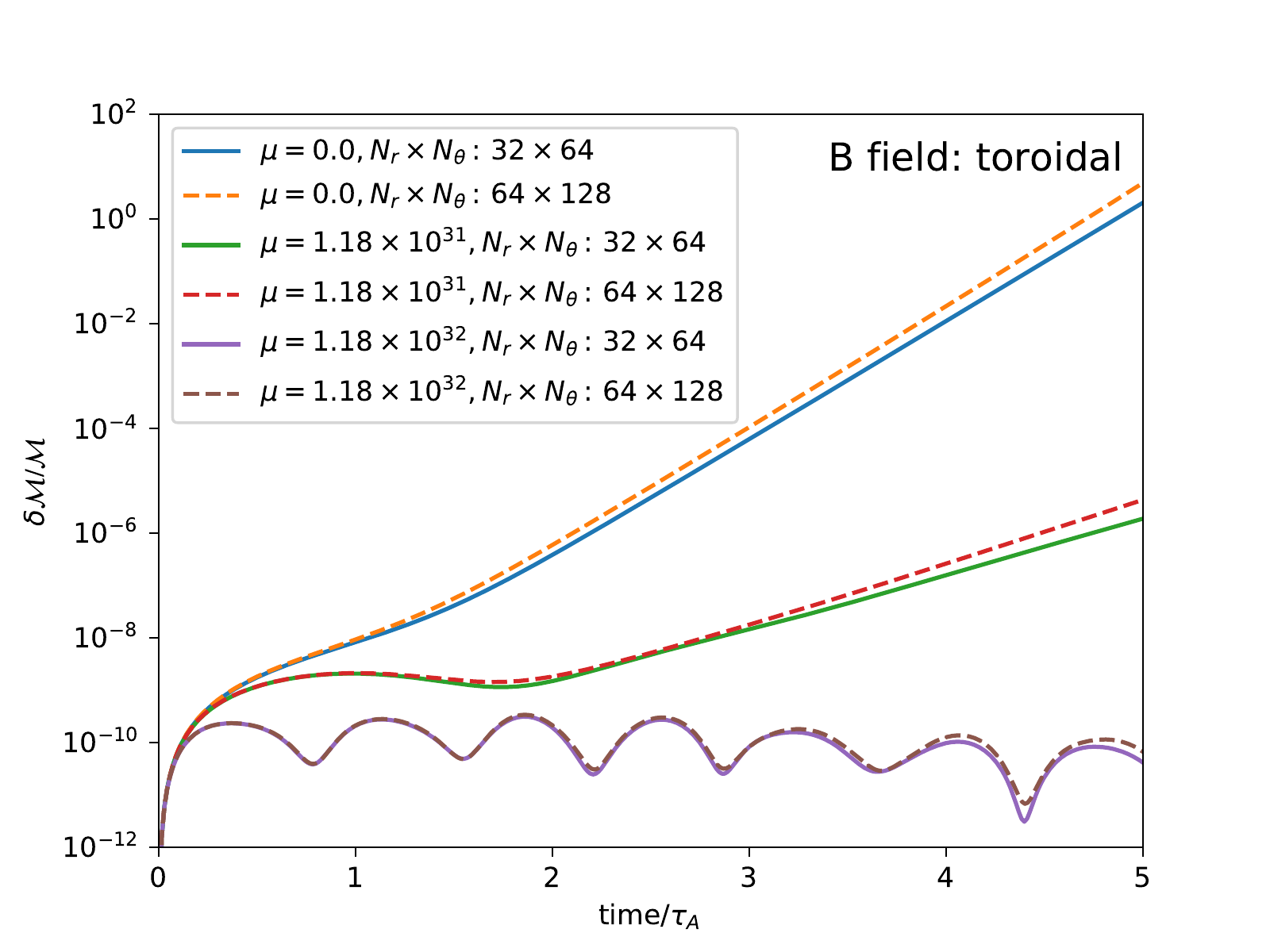}%{fig2Bevolution_with_shear.pdf}
	\includegraphics[width=\columnwidth]{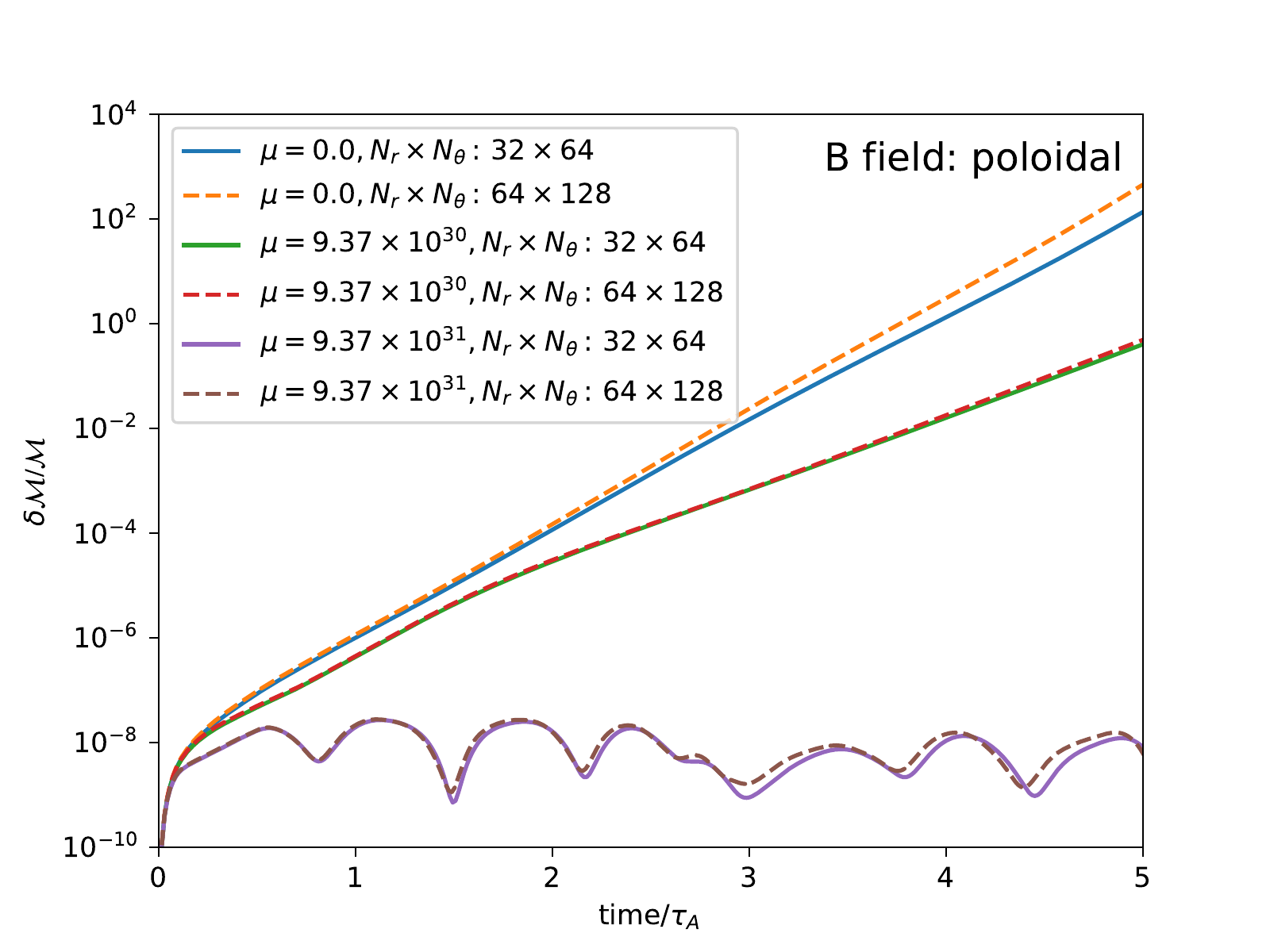}
    \caption{\textit{Left}: The time evolution of perturbed magnetic energy ($\delta\mathcal{M}$) corresponding to an $m=1~$mode in a polytropic star with different values of uniform shear modulus throughout its interior. The neutron star has a pure toroidal magnetic field and the energy ratio between magnetic to gravitational is $|\mathcal{M/W}|=8\times10^{-3}$. As the strength of the shear modulus ($\mu$ in unit $\rm{Joule\cdot m^{-3}}$) increases, the characteristic growth of the instability reduces. The consistency of the numerical solution for different grid resolutions is also shown.
    \textit{Right}: A similar characteristic is shown in the time evolution of perturbed magnetic energy with an $m=2~$mode where the equilibrium star has a pure poloidal magnetic field.
    }
    \label{fig2:evolution_in_shear}
\end{figure*}
\begin{figure*}%-------------------------------------------------------------------
	\includegraphics[width=2\columnwidth]{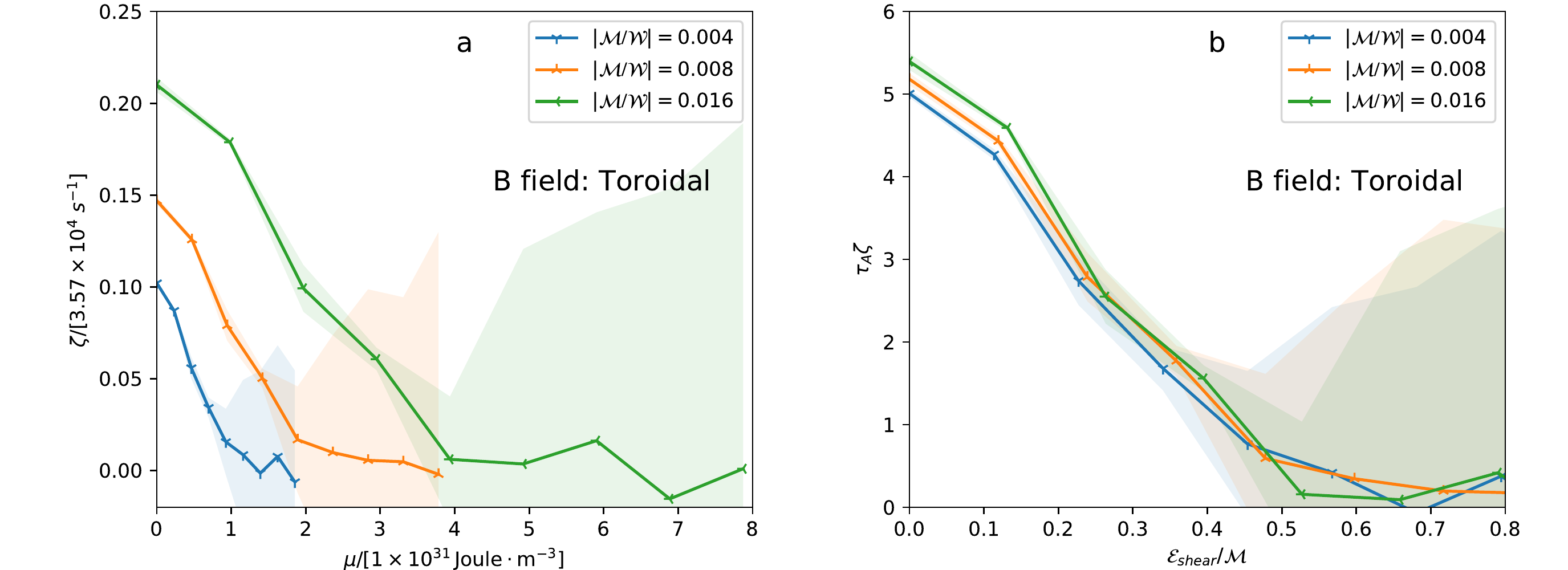}%{fig2rateVSmu.pdf}
    \caption{a): Shear modulus ($\mu$) dependent instability growth rate ($\zeta$) in neutron stars with pure toroidal magnetic field for three different magnetic energies. The instability growth rate reduces for a higher value of the shear modulus. The shaded regions connected to each curve represent the standard deviation of the growth rate about the arithmetic mean (i.e. $\zeta^m\pm\zeta^\sigma$) of time series data in the period [$2\tau_A,6\tau_A$]. The temporal behaviour of the perturbed components captures oscillatory phenomena and the standard deviation increases when the oscillation amplitude grows. For fixed magnetic energy, the minimum value $\mu_{crit}$ suppress the magnetic instability is obtained from the condition $\zeta^m\rightarrow 0$. b) Different magnetic energy configurations exhibit a self-similar feature when the instability growth rate ($\zeta$) and the corresponding shear energy ($\mathcal{E}_{shear}$) are scaled using magnetic parameters $\tau_A^{-1}$ and $\mathcal{M}$, respectively. A small mismatch in the vertical scale may be due to the simplified expression of $\tau_A$, obtained from the global average values.
    }
    \label{fig3:rateVSmu}
\end{figure*}

\begin{figure*}%-------------------------------------------------------------------
	\includegraphics[width=2\columnwidth]{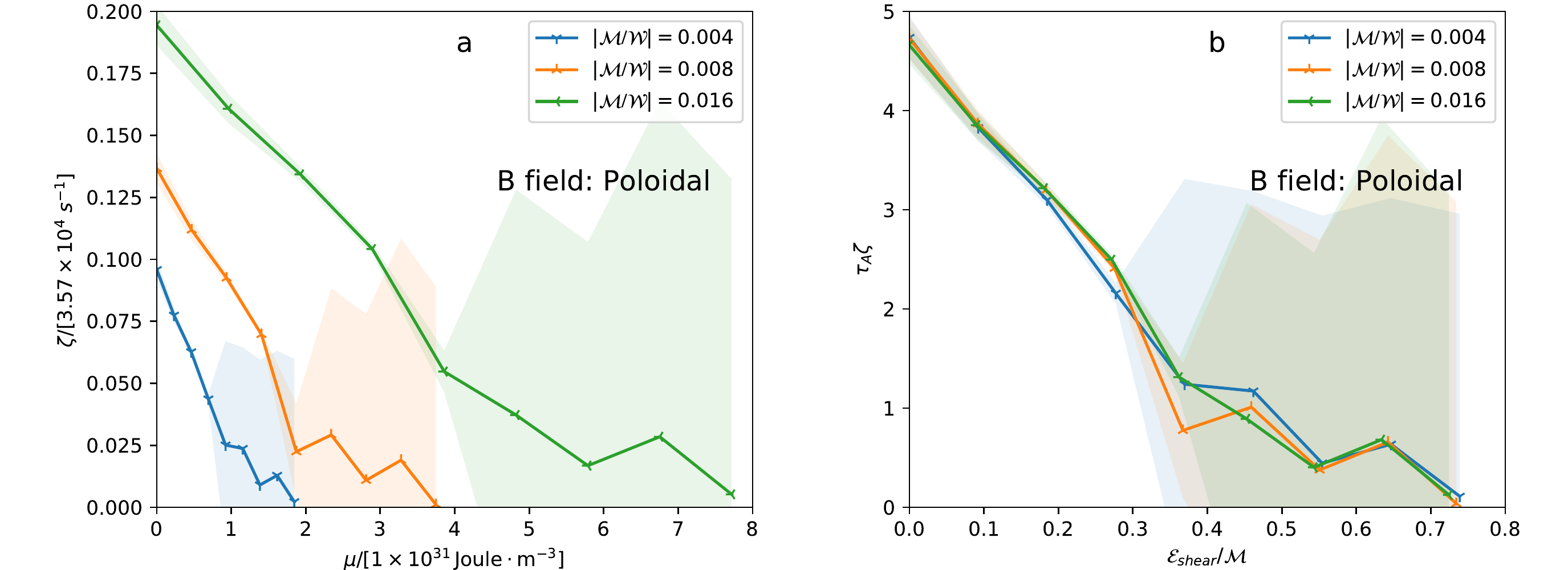}%{fig4PrateVSmu.pdf}
    \caption{Same as Figure~\ref{fig3:rateVSmu} but for a perturbed pure poloidal magnetic field configuration. Here the instability growth rate also reduces as the shear modulus ($\mu$) increases. 
    }
    \label{fig4P:rateVSmu}
\end{figure*}

%>---------------------------------------------------------------

In the previous section (section~\ref{S3}), we demonstrated that an axisymmetric fluid star, without any shear modulus, shows magnetic Tayler instability for a purely poloidal or toroidal magnetic field geometry. The linear order time evolution shows exponential growth in the perturbed values as characteristic of the Tayler instability. The growth rate is related to the corresponding Alfv\'en-crossing time as reported in other papers \citep{Lander+Jones2011a,Lander+Jones2011b,Bera+Bhattacharya2017}. In this section, we assume a star with a uniform shear modulus throughout the interior. This may not be a realistic assumption but it is a natural starting point. 

A forward time evolution with a different value of uniform shear modulus all over the star indicates that as the strength of shear modulus increases, it reduces the growth of the instability (Figure~\ref{fig2:evolution_in_shear}). For a very high value of the shear modulus, the configurations only oscillate about the equilibrium. Both the poloidal and toroidal magnetic field geometries show similar behaviour. We evolve the equations with different grid resolutions to test the accurate implication of the numerical setup. From the Figure~\ref{fig2:evolution_in_shear} we find the results are almost independent of the grid resolutions used for the calculation. 

The instability growth rate gradually decreases as the strength of shear modulus increases (Figure~\ref{fig3:rateVSmu}~\&~\ref{fig4P:rateVSmu}). After the initial settling time of about $\tau_A$, the instability growth-rate attends a steady value when the value of shear modulus is low. An increased shear modulus weakens the instability. The time evolution shows oscillatory features when the shear modulus is very large. The configuration oscillates about the equilibrium and hence, the instantaneous growth rate changes as time passes. To capture this time-dependent growth rate feature, we calculate an arithmetic mean of the growth rate ($\zeta^m$) over the period [2$\tau_A$,~6$\tau_A$]. The enhancement of the oscillation amplitude increases the standard deviation of the growth rate ($\zeta^\sigma$) about the mean $\zeta^m$ (Figure~\ref{fig3:rateVSmu}a,\ref{fig4P:rateVSmu}a). The average growth rate of instability over the Alfv\'en time is $\zeta=\zeta^m$, whereas the instantaneous growth rate at any moment is most probable to lie within the shaded region ($\zeta^m\pm\zeta^\sigma$). The value of critical shear modulus ($\mu_{crit}$) is estimated from the condition when the instability growth rate tends to vanish (i.e. $\zeta^m\rightarrow0$). We find that the value of $\mu_{crit}$ is about $2\times10^{31}~\rm{Joule\cdot m^{-3}}$ for both pure toroidal and pure poloidal magnetic field configurations with $|\mathcal{M/W}|=8\times10^{-3}$. Similarly, we have found the characteristics of the instability growth rate for the configurations with different magnetic energy $|\mathcal{M/W}|=4\times10^{-3}~\&~16\times10^{-3}$. Figures~\ref{fig3:rateVSmu}a~\&~\ref{fig4P:rateVSmu}a show that as the magnetic energy of the star increases the required shear modulus becomes larger to prevent the magnetic instability, i.e. $\mu_{crit}$ is higher for a star with larger $\mathcal{M}$.

In a fluid star, as the magnetic field strength increases the instability growth rate increases. The shear modulus dependent instability growth rates follow a similar trend for different magnetic energies. If the growth rate and shear modulus are scaled using $\tau_A$ and $|\mathcal{M/W}|$ respectively, the different curves relating magnetic energy display a form of  self-similarity, as shown in  Figure~\ref{fig3:rateVSmu}b and \ref{fig4P:rateVSmu}b. In addition to the similarity feature, we can get some idea about the value of $\kappa$ (see equation (\ref{eq_instability_shear})) required to have a stable oscillation. Figure~\ref{fig3:rateVSmu}b and \ref{fig4P:rateVSmu}b indicate the value of $\kappa$ are about $0.5$ and $0.4$ for pure toroidal and pure poloidal magnetic field geometries respectively.

A barotropic star with axisymmetric mixed poloidal and toroidal field, obtained from self-consistent field studies, also exhibits Tayler instability as the configurations are mainly dominated with a poloidal fraction \citep{Lander+Jones2012,Bera+Bhattacharya2017}. Hence, we expect a similar effect due to the shear modulus on magnetic field evolution in such field geometries.

%---------------------------------------------------------------<

\subsection{Evolution of stellar magnetic field with crustal shear} \label{sS4_2}

%>---------------------------------------------------------------

The results presented in the previous section are obtained for a star with a uniform shear modulus throughout the interior. To restrict the effect to the region of the stellar crust, we consider a radial profile of the shear modulus which gradually increases from the centre. For simplicity, we assume a smoothly varying radial profile of the shear modulus to model the localised effects of the elastic crust. This avoids complexities related to a sharp gradient of elasticity and related issues associated with the implementation of an inner boundary within the computational domain. This model is not particularly realistic but provides qualitative insight.

\begin{figure*}%-------------------------------------------------------------------
	\includegraphics[width=2\columnwidth]{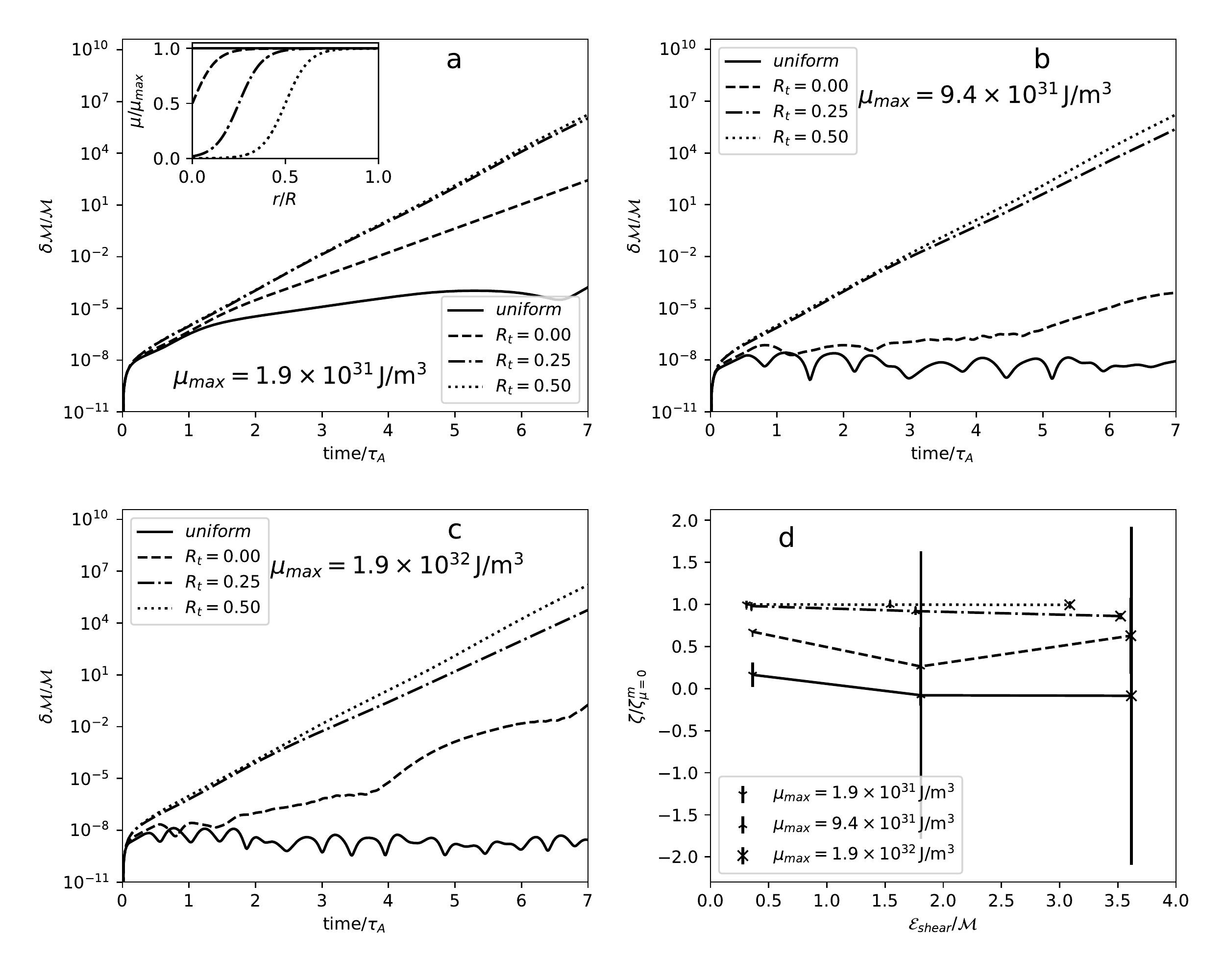}%{radial_mu_Bevolution.pdf}
    \caption{The time evolution of perturbed magnetic energy of an equilibrium neutron star with a poloidal magnetic field and energy ratio $|\mathcal{M/W}|=0.008$ for different radial profiles (inset figure) of the shear modulus. 
    a) The perturbed magnetic energy in the $m=2$ mode grows as time passes and indicates the presence of instability for a radially uniform shear modulus ( $1.9\times10^{31}~\rm{Joule\cdot m^{-3}}$). For a different radial profile of the shear modulus within the star, the perturbation grows at a higher rate when the shear modulus is smaller close to the unstable region ($r/R\sim0.6$, Figure~\ref{fig1:static}b). The radial profiles of the shear modulus are shown in the figure inset. b) and c) are same as figure a) but for $\mu_{max}$ = $9.4\times10^{31}~\rm{Joule\cdot m^{-3}}$ and $1.9\times10^{32}~\rm{Joule\cdot m^{-3}}$ respectively. The instability reduces significantly for the uniform cases but the growth rate depends on $R_t$, i.e. the radial profile of shear modulus. d) The calculated growth rates ($\zeta^m\pm\zeta^\sigma$) in the time range [2$\tau_A$, 6$\tau_A$] of Figures~\ref{fig5}~a,~b~\&~c  and the corresponding volume integrated shear energy ($\mathcal{E}_{shear}$) are shown here. The instability growth rate almost achieves the maximum, corresponding to the no-shear condition ($\mu=0$), for the radial profile $R_t=0.5$, irrespective of the value of $\mathcal{E}_{shear}$. 
    }
    \label{fig5}
\end{figure*}

We consider a radial profile of shear modulus with the following functional form:
\begin{align}\label{radial_mu_profile}
\mu(r)=\frac{\mu_{max}}{2}\left(1+\tanh\frac{r/R-R_t}{\delta R_t}\right).
\end{align}
This form ensures $\mu\simeq\mu_{max}$ as $r\to R$, and the value of $\mu$ gradually decreases from the surface towards the core. Here, $R$ is the equatorial radius of neutron star and $R_t$ represents the radial location (in unit of $R$) where $\mu=\mu_{max}/2$. $\delta R_t$ is a parameter which decides the effective radial width over which the gradual transition of shear modulus happens near $R_t$. We keep a fixed value of $\delta R_t=0.125$. A set of radial profiles of shear modulus are shown in the inset of Figure~\ref{fig5}a. The radial profiles corresponds to $R_t=0$, $R_t=1/4$ and $R_t=1/2$. A uniform radial function is also shown for the comparison. Very close to the stellar surface $r=R$, the shear modulus is expected to drop sharply as the matter density vanishes. This is not captured in equation~\ref{radial_mu_profile}. The instability features for some other radial profiles are given in Appendix~\ref{appendix_B}.

\begin{figure*}%-------------------------------------------------------------------
	\includegraphics[width=2\columnwidth]{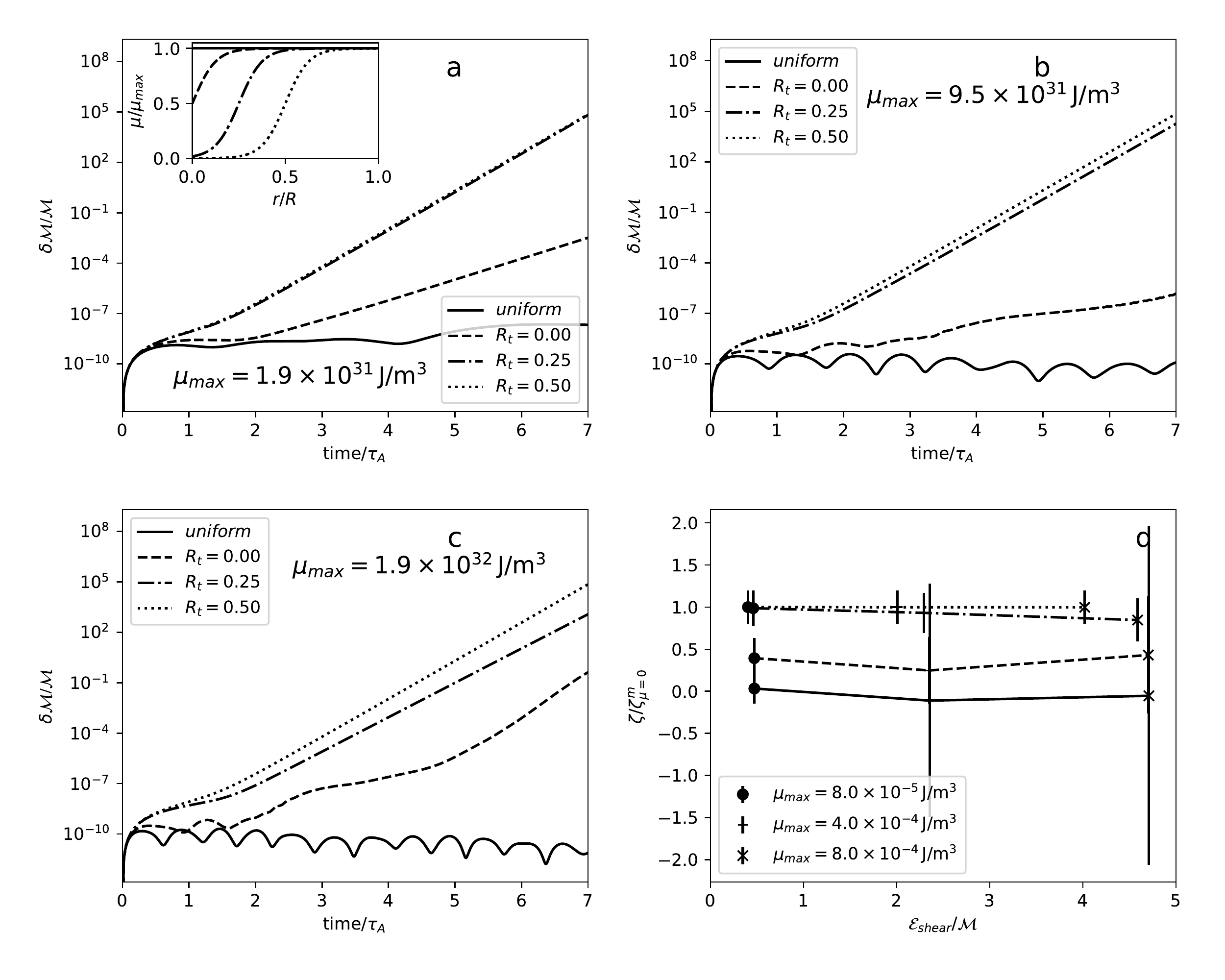}
    \caption{Same as Figure~\ref{fig5}, but the perturbation characteristics are due to an $m=1$ mode for the equilibrium configuration of Figure~\ref{fig1:static}a with pure toroidal magnetic field.
    }
    \label{fig6}
\end{figure*}

For a poloidal magnetic field configuration with $|\mathcal{M/W}|=0.008$ (Figure~\ref{fig1:static}b), we choose $\mu_{max}\gtrsim\mu_{crit}$, and find the instability characteristics for different radial profiles of shear modulus. This pure poloidal magnetic field configuration with a uniform shear modulus suggests a critical value near about $\mu_{crit}=1.9\times10^{31}~\rm{Joule\cdot m^{-3}}$ as $\zeta$ tends to vanish (Figure~\ref{fig4P:rateVSmu}). The stars with a uniform shear modulus throughout its interior such that $\mu_{max}>\mu_{crit}$ show stable oscillation when it is perturbed (Figure~\ref{fig2:evolution_in_shear}b). However, a reduced shear modulus in the core region causes  the system to exhibit magnetic instability (Figure~\ref{fig5}a). The configurations with an insignificant shear modulus at the core also show a similar behaviour even for a higher value of $\mu_{max}$ (Figure~\ref{fig5}b,c). The radial profile of shear modulus with $R_t=0.5$ shows maximum instability growth even for the case when $\mu_{max}\approx10\cdot\mu_{crit}$. The time evolution characteristics of Figure~\ref{fig5}a,b,c are summarised by calculating the instability growth rate and presented in Figure~\ref{fig5}d with their $\mathcal{E}_{shear}$ value. It shows the presence of instability is dependent on the radial profile of shear modulus within the star, and almost independent of the volume integrated shear modulus. Therefore, the value of $\kappa$ in equation~\ref{eq_instability_shear} does depend on the profile of shear modulus and the magnetic field geometry. For a pure toroidal magnetic field geometry, the radial profile of the shear modulus exhibits similar instability features (Figure~\ref{fig6}). Hence, it suggests the magnetic Tayler instability is intrinsic to the field geometry and local environment near the magnetic null region.

%---------------------------------------------------------------<

\section{Discussion and conclusion}\label{S5}

We have investigated the impact of elasticity on the magnetic Tayler instability in an idealised polytropic star. We did this by performing time evolutions of the perturbed magneto-hydrodynamic equations,  starting from an axisymmetric equilibrium configuration. We found that the growth rate of the instability in a magnetic star reduces when the shear modulus is large enough to make an impact based on its strength and spatial distribution.   We now discuss a few points relevant to applying our results to real neutron stars:

\textit{Magnetic field geometry:} Observational properties of a neutron star indicate the presence of a strong multi-polar surface magnetic field components \citep{Bilous+2019}. But there is no prior direct evidence on the interior field geometry. To make the calculation simple we considered axisymmetric stellar configurations with pure poloidal and pure toroidal magnetic fields which cover almost the entire star (Figure~\ref{fig1:static}). Our results demonstrate that the presence of an elastic environment suppresses the Tayler instability of these configurations. In an overall scenario, the dynamical properties of a localised magnetic field might follow qualitatively the characteristics of the extended field geometries (e.g. the threshold criteria of equation~\ref{eq_instability_shear}).

\textit{Tayler instability and crustal shear}: The electrostatic interactions between crystallized ions in the crust of a neutron star generate the elastic properties. The value of shear modulus and its non-isotropic nature depends on the crystal parameters and the local environment, e.g. temperature, magnetic field \citep{Sato1979}. A typical value of the shear modulus at the neutron star crust is about $\mu_{eff}\sim10^{29}~\rm{Joule\cdot m^{-3}}$.\footnote{The effective shear modulus in the neutron star crust can be estimated as $\mu_{eff}\sim 0.1*n(Ze)^2/a$ \citep{Horowitz+Hughto2008arXiv0812.2650H}. Here, $n\sim(4/3\pi a^3)^{-1}$ is the ion number density and $a$ is the radius of a Wigner-Seitz cell containing one single ion of charge $Ze$. For the crust with $^{56}\rm{Fe}_{26}$ ion in a crystal with $a>2.2\times10^{-16}~\rm{cm}$ effective shear modulus becomes $\mu_{eff}\sim 3.7 \times10^{29}~\rm{Joule\cdot m^{-3}}$.} The presence of a magnetic field may influence the crust structure in a neutron star \citep{Chamel+2012,Franzon+2017}. It may also introduce a local magnetic field dependent non-isotropic shear modulus \citep{Baiko+Kozhberov2017}. Assuming these as higher-order corrections, we neglect these details in our study to keep the calculation simple.

\begin{figure*}%-------------------------------------------------------------------
	\includegraphics[width=2\columnwidth]{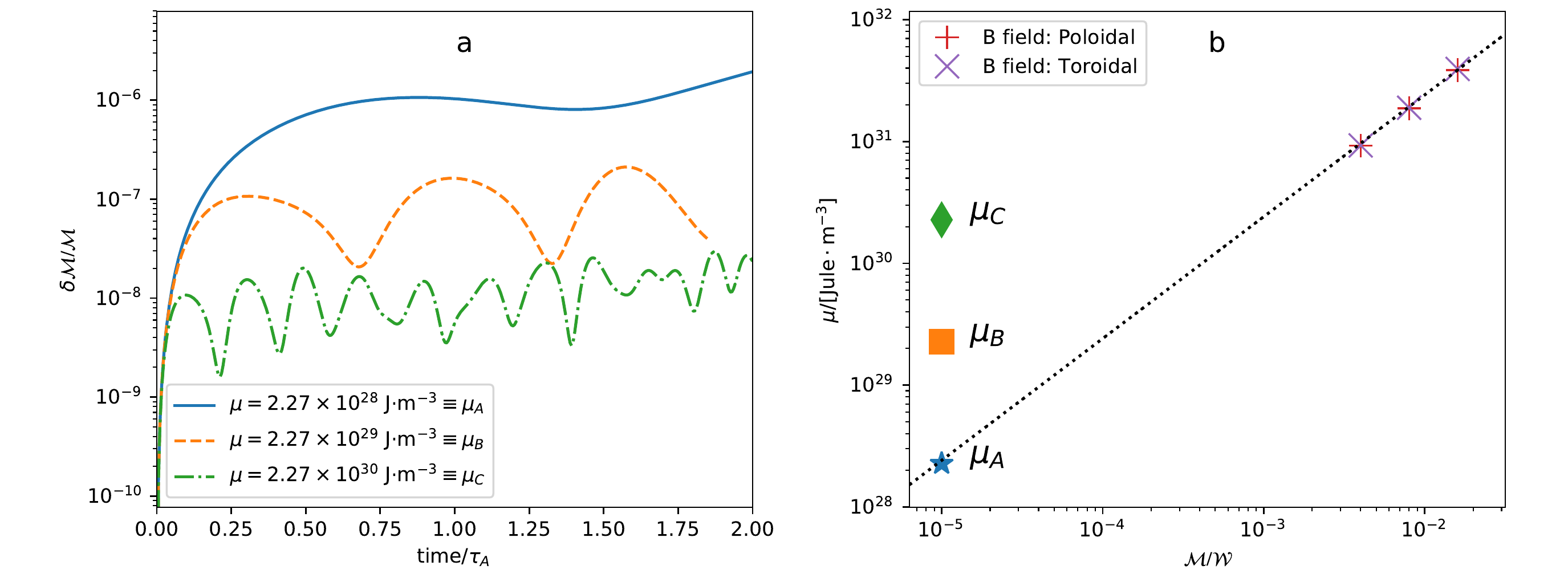}
    \caption{a) Same as Figure~\ref{fig2:evolution_in_shear}, but for a equilibrium star with a volume-averaged toroidal magnetic field $3\times10^{11}$~T ($|\mathcal{M/W}|=10^{-5}$). In this case, the Alfv\'en time ($\tau_A\sim0.03$~s) is about 28 times of the configuration presented in Figure~\ref{fig1:static}a (and hence it leads to a very long computation time). A uniform shear modulus of strength above $10^{29}~\rm{Joule\cdot m^{-3}}$ within the star can exhibit stable oscillation. b) The linear extrapolation of the critical shear modulus ($\mu_{crit}$) values from Figures~\ref{fig3:rateVSmu}~\&~\ref{fig4P:rateVSmu} indicates that the uniform shear modulus $\mu_A=2.3\times10^{28}~\rm{Joule\cdot m^{-3}}$ is close to the critical value ($\mu_{crit}$) of this configuration to exhibit stability against perturbation. 
    }
    \label{fig7}
\end{figure*}

In section~\ref{sS4_1}, we considered stars with a uniform shear modulus throughout the entire stellar volume.  We found that a shear modulus of about $2\times10^{31}~\rm{Joule\cdot m^{-3}}$ can prevent magnetic Tayler instability originating from an average magnetic field strength about $10^{13}$~T (Figure~\ref{fig2:evolution_in_shear}). The evolution of the perturbed stellar variables is observable in the sound crossing time ($\tau_s$), but the appearance of the Tayler instability can be identifiable after a few Alfve\'n crossing time ($\tau_A$). The presence of a strong magnetic field in these configurations ensured a shorter evolution time (as $\tau_A\propto\frac{1}{\bar{B}_0}$) to exhibit the magnetic field strength dependent characteristics. The short evolution time (of duration a few $\tau_A$) helped in the numerical computation in two ways: i) smaller errors even for lower grid resolution, and ii) reduced computation hours where the minimum time step is connected to the grid size and $\tau_s$. We can utilize the general characteristics of Figures~\ref{fig3:rateVSmu}~\&~\ref{fig4P:rateVSmu} to estimate the critical value of the shear modulus at a lower magnetic field strength, assuming a linear extrapolation. It suggests a star with a uniform shear modulus equal to a typical crustal value ($\sim10^{29}~\rm{Joule\cdot m^{-3}}$) can sustain a magnetic field of strength up to about $10^{11}$~T (Figure~\ref{fig7}). This field strength is higher than the inferred maximum magnetic field values ($\sim10^{8-10}$~T) at the surface of neutron stars \citep{Konar2017,Petri2019}. We can therefore say that a star with a non-zero shear modulus throughout its entire volume, equal to the typically estimated crustal value, would be capable of stabilising a neutron star with a realistic magnetic field strength.

However, the elastic component of a realistic neutron star will probably be confined to the crust.  In this case our findings of section~\ref{sS4_2} indicate that the elastic crust may not be able to stabilise the star, even if its shear modulus is increased beyond the values normally considered for crusts. Here we can conclude that the axisymmetric barotropic magnetic neutron stars with an elastic crust at the outer surface and interior magnetic field are unstable.

In summary, we have studied the instability characteristics of axisymmetric magnetic field configurations. In a fluid star, the instability growth rate is directly dependent on the Alfv\'en crossing time. The main conclusions are:  
\begin{itemize}
    \item[i)] The inclusion of elasticity with uniform shear modulus, present throughout the entire stellar volume, reduces the instability growth rate. The instability may even disappear if the shear modulus exceeds a critical value. 
    \item[ii)] An axisymmetric magnetic star with an extended magnetic field and an effective shear modulus confined to an outer shell does show a Tayler instability. Therefore a localised shear modulus may not always have a global impact on the magnetic Tayler instability. 
\end{itemize}

As it appears that the inclusion of elasticity does not resolve the stability issues associated with a generic neutron star magnetic field, we need to consider alternative explanations for the simple fact that real neutron stars support strong---obviously long-term stable---fields. Inevitably, we need to make the stellar model more complex. In order to focus on the elastic properties, we simplified many aspects of the problem. We explored liner-order perturbations of an axisymmetric magnetic and barotropic star close to equilibrium. A more realistic description would include the expected composition and state of matter and should account for evolutionary aspects. After all, a neutron star is a dynamical system, the properties of which evolves as it matures on time scales ranging from a fraction of second to giga-years. A realistic evolution involves changes in temperature, nuclear composition, magnetic field geometry \citep{Goldreich+Reisenegger1992,Gourgouliatos+Cumming2014}, rotation speed \citep{Pons+Vigano2019} etc. The different aspects are not independent and their impact on the magnetic field configuration, which may appear subtle, may be crucial. Therefore, to address the nature of the neutron star magnetic field one ought to consider a wider range of physical conditions, e.g. i) stable (composition) stratification, ii) multipolar magnetic fields, iii) a gradually evolving system, iv) the presence of matter in the superfluid/superconductor state \citep{Sauls1989,Glampedakis+2011,Graber+2015} etc. These are all interesting aspects which we hope to consider in future work. 

\section*{Acknowledgements}
PB acknowledges the use of the IRIDIS High Performance Computing Facility, and associated support services at the University of Southampton, in the completion of this work. We acknowledge financial support from STFC via Grant No. ST/R00045X/1. We thank the anonymous referees for constructive comments.

\section*{Data availability}
The data underlying this article will be shared on reasonable request to the corresponding author.

%%%%%%%%%%%%%%%%%%%%%%%%%%%%%%%%%%%%%%%%%%%%%%%%%%



\bibliographystyle{mnras}
\bibliography{NS_perturb} % if your bibtex file is called example.bib

%\begin{thebibliography}{99}
%\end{thebibliography}

%%%%%%%%%%%%%%%%%%%%%%%%%%%%%%%%%%%%%%%%%%%%%%%%%%

%%%%%%%%%%%%%%%%% APPENDICES %%%%%%%%%%%%%%%%%%%%%

\appendix

\section{shear terms}\label{shear_terms}
 In a spherical polar coordinate system, the directional differentiation and the connection coefficients $\Gamma_{ijk}$ are:
\begin{align}
    \xi_{i,r}=\frac{\partial\xi_i}{\partial r}, ~\xi_{i,\theta}=\frac{1}{r}\frac{\partial\xi_i}{\partial \theta}, ~\xi_{i,\phi}=\frac{1}{r\sin\theta}\frac{\partial\xi_i}{\partial \phi};\\
    \Gamma_{\theta r \theta} = \Gamma_{\phi r \phi} = -\Gamma_{r\theta\theta}= -\Gamma_{r\phi\phi}=\frac{1}{r},~ \Gamma_{\phi\theta\phi}=-\Gamma_{\theta\phi\phi}=\frac{\cot\theta}{r}.
\end{align}
Now, $S_{ik} \equiv \xi_{i;k}=\xi_{i,k}+\Gamma_{ijk}\xi_j$. Hence, the shear terms in a spherical polar coordinate appear in the following form:
\begin{align}
    \Sigma_{rr} &= \frac{2}{3}\frac{\partial\xi_r}{\partial r} - \frac{2}{3r}\xi_r - \frac{\cot\theta}{3r}\xi_\theta - \frac{1}{3r}\frac{\partial\xi_\theta}{\partial\theta} - \frac{1}{3r\sin\theta}\frac{\partial\xi_\phi}{\partial\phi};\\
    \Sigma_{\theta\theta} &= \frac{2}{3r}\frac{\partial\xi_\theta}{\partial\theta} + \frac{\xi_r}{3r} -\frac{1}{3}\frac{\partial\xi_r}{\partial r} - \frac{\cot\theta}{3r}\xi_\theta - \frac{1}{3r\sin\theta}\frac{\partial\xi_\phi}{\partial\phi};\\
    \Sigma_{r\theta} &= \Sigma_{\theta r} = \frac{1}{2}\frac{\partial\xi_\theta}{\partial r} - \frac{\xi_\theta}{2r} + \frac{1}{2r}\frac{\partial\xi_r}{\partial\theta};\\
    \Sigma_{\phi\phi} &= \frac{2}{3r\sin\theta}\frac{\partial\xi_\phi}{\partial\phi} + \frac{2}{3}\frac{\cot\theta}{r}\xi_\theta + \frac{\xi_r}{3r} - \frac{1}{3}\frac{\partial\xi_r}{\partial r} - \frac{1}{3r}\frac{\partial\xi_\theta}{\partial\theta};\\
    \Sigma_{r\phi} &= \Sigma_{\phi r} = \frac{1}{2r\sin\theta}\frac{\partial\xi_r}{\partial\phi}+\frac{1}{2}\frac{\partial\xi_\phi}{\partial r} - \frac{1}{2r}\xi_\phi;\\
    \Sigma_{\theta\phi} &= \Sigma_{\phi\theta} = \frac{1}{2r}\frac{\partial\xi_\phi}{\partial\theta} + \frac{1}{2r\sin\theta}\frac{\partial\xi_\theta}{\partial\phi} - \frac{\cot\theta}{2r}\xi_\phi.
\end{align}

The stress due to trace-less strain is obtained using the shear modulus $\mu$,
\begin{align}
    \mathbf{T}= - 2\mu\boldsymbol{\Sigma}.
\end{align}
Hence, the force term due to shear (equation~\ref{F_shear}) can be expressed as,
\begin{align}
    \boldsymbol{\mathcal{F}}_{shear} = -\nabla\cdot\textbf{T} = 2\mu\nabla\cdot\boldsymbol{\Sigma}+2\nabla\mu\cdot\boldsymbol{\Sigma}.
\end{align}
In a medium with a uniform shear modulus ($\nabla\mu=0$), the spherical polar components of $\boldsymbol{\mathcal{F}}_{shear}$ appear in the following form:
\begin{align}\label{F_shear_phi}
    \mathcal{F}_\phi|_{shear}/{2\mu} &= \Sigma_{\phi k;k} = \Sigma_{\phi r;r} + \Sigma_{\phi\theta;\theta} + \Sigma_{\phi\phi;\phi}\nonumber\\
    &= \left(\frac{\partial\Sigma_{\phi r}}{\partial r} + \Gamma_{\phi k r}\Sigma_{kr} + \Gamma_{rkr}\Sigma_{\phi k}\right) \nonumber\\ &~~~~+ \left(\frac{1}{r}\frac{\partial\Sigma_{\phi \theta}}{\partial \theta} + \Gamma_{\phi k \theta}\Sigma_{k\theta} + \Gamma_{\theta k\theta}\Sigma_{\phi k}\right) \nonumber\\ &~~~~+ \left(\frac{1}{r\sin\theta}\frac{\partial\Sigma_{\phi \phi}}{\partial \phi} + \Gamma_{\phi k \phi}\Sigma_{k\phi} + \Gamma_{\phi k\phi}\Sigma_{\phi k}\right)\nonumber\\
    &= \frac{\partial \Sigma_{\phi r}}{\partial r} + \frac{1}{r}\frac{\partial \Sigma_{\phi\theta}}{\partial \theta} + \frac{1}{r\sin\theta}\frac{\partial\Sigma_{\phi\phi}}{\partial \phi} \nonumber\\ &~~~~+\frac{3}{r}\Sigma_{\phi r} + \frac{2\cot\theta}{r}\Sigma_{\theta\phi};
\end{align}

\begin{align}\label{F_shear_theta}
    \mathcal{F}_\theta|_{shear}/{2\mu} &= \Sigma_{\theta k;k} = \Sigma_{\theta r;r} + \Sigma_{\theta\theta;\theta} + \Sigma_{\theta\phi;\phi}\nonumber\\
    &= \left(\frac{\partial\Sigma_{\theta r}}{\partial r} + \Gamma_{\theta k r}\Sigma_{kr} + \Gamma_{rkr}\Sigma_{\theta k}\right) \nonumber\\ &~~~~+ \left(\frac{1}{r}\frac{\partial\Sigma_{\theta \theta}}{\partial \theta} + \Gamma_{\theta k \theta}\Sigma_{k\theta} + \Gamma_{\theta k\theta}\Sigma_{\theta k}\right) \nonumber\\ &~~~~+ \left(\frac{1}{r\sin\theta}\frac{\partial\Sigma_{\theta \phi}}{\partial \phi} + \Gamma_{\theta k \phi}\Sigma_{k\phi} + \Gamma_{\phi k\phi}\Sigma_{\theta k}\right)\nonumber\\
    &= \frac{\partial \Sigma_{\theta r}}{\partial r} + \frac{1}{r}\frac{\partial \Sigma_{\theta\theta}}{\partial \theta} + \frac{1}{r\sin\theta}\frac{\partial\Sigma_{\theta\phi}}{\partial \phi} + \frac{2}{r}\Sigma_{r\theta}\nonumber\\ &~~~~ -\frac{\cot\theta}{r}\Sigma_{\phi \phi} + \frac{\cot\theta}{r}\Sigma_{\theta\theta};
\end{align}

\begin{align}\label{F_shear_r}
    \mathcal{F}_r|_{shear}/{2\mu} &= \Sigma_{r k;k} = \Sigma_{r r;r} + \Sigma_{r\theta;\theta} + \Sigma_{r\phi;\phi}\nonumber\\
    &= \left(\frac{\partial\Sigma_{r r}}{\partial r} + \Gamma_{r k r}\Sigma_{kr} + \Gamma_{rkr}\Sigma_{r k}\right) \nonumber\\ &~~~~+ \left(\frac{1}{r}\frac{\partial\Sigma_{r \theta}}{\partial \theta} + \Gamma_{r k \theta}\Sigma_{k\theta} + \Gamma_{\theta k\theta}\Sigma_{r k}\right) \nonumber\\ &~~~~+ \left(\frac{1}{r\sin\theta}\frac{\partial\Sigma_{r \phi}}{\partial \phi} + \Gamma_{r k \phi}\Sigma_{k\phi} + \Gamma_{\phi k\phi}\Sigma_{r k}\right)\nonumber\\
    &= \frac{\partial \Sigma_{r r}}{\partial r} + \frac{1}{r}\frac{\partial \Sigma_{r\theta}}{\partial \theta} + \frac{1}{r\sin\theta}\frac{\partial\Sigma_{r\phi}}{\partial \phi} - \frac{1}{r}\Sigma_{\theta\theta}\nonumber\\ &~~~~+\frac{2}{r}\Sigma_{r r} - \frac{1}{r}\Sigma_{\phi\phi} + \frac{\cot\theta}{r}\Sigma_{r\theta}.
\end{align}

For a $r$-dependent shear modulus $\mu(r)$, the components $\mathcal{F}_\phi|_{shear}$ and $\mathcal{F}_\theta|_{shear}$ remain as these are in equations~\ref{F_shear_phi}~\&~\ref{F_shear_theta} respectively but $\mathcal{F}_r|_{shear}$ includes extra terms with the factor $\frac{\partial \mu}{\partial r}$ in its expression:
\begin{align}
    \mathcal{F}_r|_{shear}= 2\mu\Sigma_{r k;k} + 2\frac{\partial \mu}{\partial r}(\Sigma_{rr}+\Sigma_{r \theta}+\Sigma_{r\phi}).
\end{align}

\section{shear modulus in the interior}\label{appendix_B}
\begin{figure*}%-------------------------------------------------------------------
	\includegraphics[width=2\columnwidth]{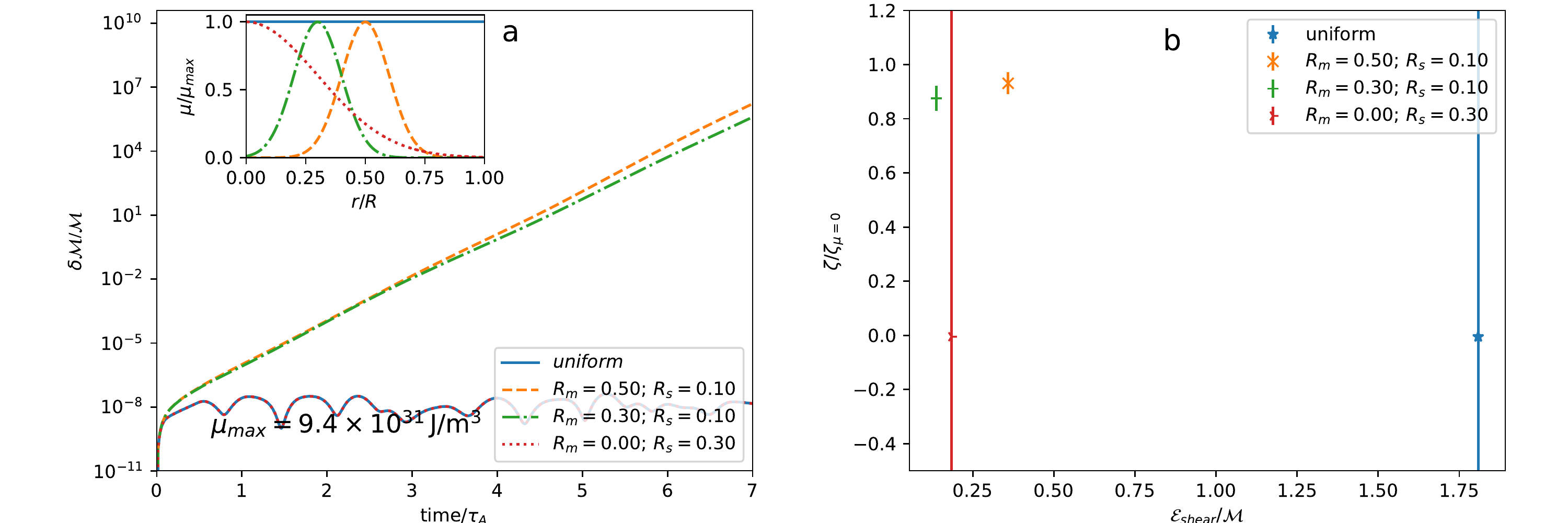}
    \caption{a) Time evolution of the perturbed variables for a poloidal magnetic field configuration of Figure~\ref{fig1:static}b with different Gaussian radial profile of shear modulus $\mu(r)=\mu_{max}\exp\left[-\frac{(r/R-R_m)^2}{2R_s^2}\right]$. b) The instability growth rates ($\zeta^m\pm\zeta^\sigma$) depends on the radial profile of the shear modulus. % from different shear profiles indicate that the value of shear near $r=0$ decides the instability characteristics. 
    }
    \label{fig8}
\end{figure*}

\begin{figure*}%-------------------------------------------------------------------
	\includegraphics[width=2\columnwidth]{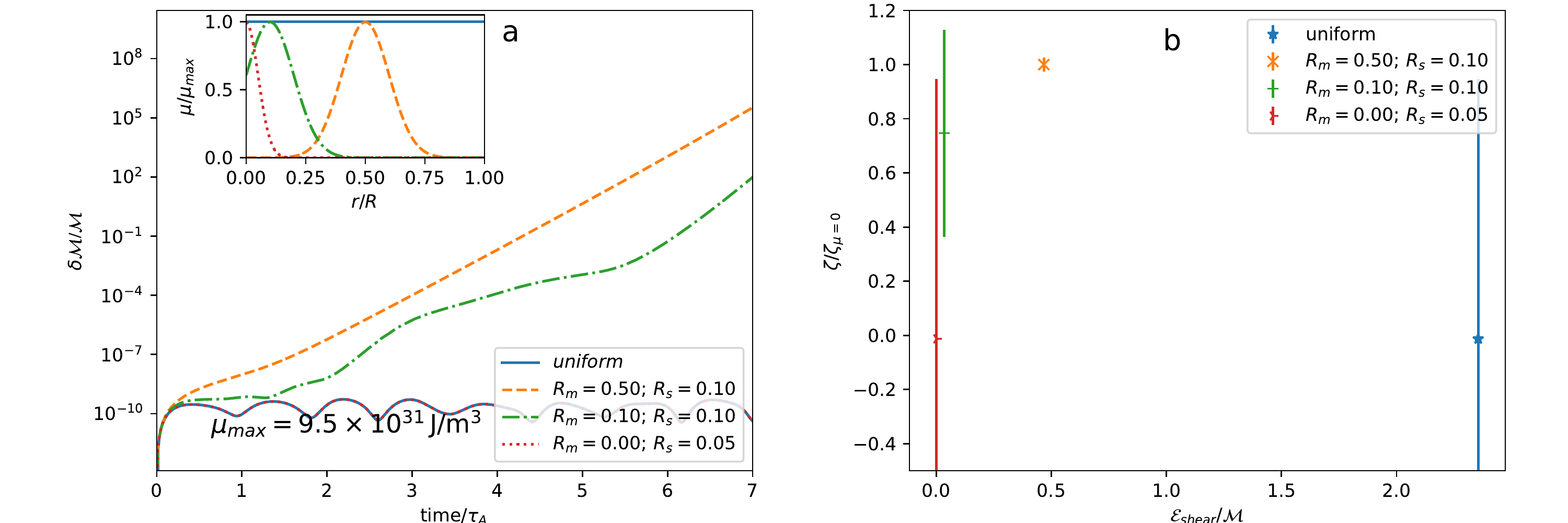}
    \caption{Same as Figure~\ref{fig8} but for a pure toroidal magnetic field configuration of Figure~\ref{fig1:static}a with different Gaussian radial profile of shear modulus.
    }
    \label{fig8b}
\end{figure*}
In section~\ref{sS4_2}, we consider a gradually increasing radial profile of shear modulus (equation~\ref{radial_mu_profile}) with a maximum near the stellar surface to connect it with the crustal shear within neutron star. Another kind of star, suppose with a crystallised matter in its interior, e.g. a white dwarf, might have a different radial profile of shear modulus. To explore these possibilities, here we also study some radial profiles of the shear modulus with a significant strength within the stellar interior compared to the surface value. This condition is not relevant in case of a neutron star with a crust but this exploration can be used to understand the local effects in the action of instability. As an example, we assume the shear modulus in the Gaussian form: 
\begin{align} 
\mu(r)=\mu_{max}\exp\left[-\frac{(r/R-R_m)^2}{2R_s^2}\right].
\end{align}
Here, $R_m$ decides the radial distance (in the unit of the stellar radius~$R$) of the maximum shear modulus $\mu_{max}$ from the centre and $R_s$ is a measure of effective radial width. 

In Figures~\ref{fig8}~\&~\ref{fig8b}, we consider the values of $\mu_{max}$ such that the uniform radial $\mu$-profiles do not exhibit magnetic instability in pure poloidal and pure toroidal magnetic field configurations respectively. The time evolution of linear order perturbation in the different shear profiles of Figures~\ref{fig8}~\&~\ref{fig8b} indicates that the value of shear modulus near the stellar core ($r=0$) influences the global growth of the magnetic instability. Reason the core region plays a significant role may be due to its very fast sound speed that transports pressure-density perturbation in the other parts of the fluid system.

%%%%%%%%%%%%%%%%%%%%%%%%%%%%%%%%%%%%%%%%%%%%%%%%%%

% Don't change these lines
\bsp	% typesetting comment
\label{lastpage}
\end{document}